# Bridging the Virtual Observatory and the GRID with the query element

G. Taffoni ([2]), E.Ambrodsi([4]),C. Vuerli([1,2]), A. Barisani([1]), R. Smareglia([1,2]), A. Volpato([3]), S. Pastore([3]), A. Baruffolo([3]), A. Ghiselli([4]), F. Pasian([1,2]), L. Benacchio([3])

([1]) INAF-Osservatorio Astronomico di Trieste

([2]) INAF-SI

([3]) INAF – Osservatorio Astronomico di Padova

([4]) INFN- CNAF

taffoni@oats.inaf.it
ambrosi@cnaf.infn.it
vuerli@oats.inaf.it
barisani@oats.inaf.it
smareglia@oats.inaf.it
volpato@ieee.org
pastore@oapd.inaf.it
baruffolo@oapd.infn.it
ghiselli@cnaf.infn.it
pasian@oats.inaf.it
benacchio@oapd.inaf.it

## ABSTRACT

The current generation of Grid infrastructures designed for production activity is strongly computing oriented and tuned on the needs of applications that requires intensive computations. Problems arise when trying to use such Grids to satisfy the sharing of data-oriented and service-oriented resources as happens in the IVOA community. We have designed, developed and implemented a Grid query element to access data source from an existing production Grid environment. We also enhanced the Grid middleware model (collective resources and sites) to manage Data Sources extending the Grid semantic. The query element and the modified grid Information System are able to connect the Grid environment to Virtual Observatory resources. A specialized query element is designed to work as Virtual Observatory resource in the Grid so than an Astronomer can access Virtual Observatory data using the IVOA standards.

## Keywords
Grid computing, dse, database, ivoa.

## 1. INTRODUCTION

The term "Grid", in the Virtual Observatory (VO) context, has mainly been used to indicate a set of interoperable services. This is rather different from the approach other scientific communities are taking, mainly based on using the grid for computational tasks.

Within this framework, it appears as extremely important to interconnect the VO and the computational Grid infrastructure, with particular attention to the development of the Theoretical VO, where the production of simulated data often require massive use of computing power.

Harmonization of the VO infrastructure and user tools with the developments being carried out within the various national and European Grid projects is highly desirable.

The Astronomical Observatories of Trieste and Padova are involved in the **Grid.it** project [6] aiming at 'Enabling platforms for high-performance computational grids orientated to scalable virtual organizations'. In this framework we explore the use of grid technologies for developing astrophysical applications. The testbed for grid applications is the Italian INFN Production Grid for Scientific Applications based on the LCG-2 (LHC Computing Grid) distribution [6].

Within Grid.it, we focused on the portability to the Grid of some applications providing access to Astronomical data resources, in particular the Second Guide Star Catalog (GSC-II) [3], the Two Mass Star Catalog and the archive of observational data from the Italian Galileo National Telescope (TNG) [10]. The final goal of this work is to allow "Grid Astronomers" to connect to Astronomical data using the standards developed by the International Virtual Observatory Alliance (IVOA) [4].

The Grid infrastructure based on LCG is an extremely stable computational and storage resource, however it does not supply an adequate model for DBMSes: it does not exist a Grid embedded system to access data sources. This is a crucial missing point of any Globus 2.4 [8] based Grid for any Astronomical application, both experimental and/or theoretical. INFN and INAF institutes are collaborating to design, develop and implement architecture to integrate Data Source and Data Source Engine in the existing grid infrastructure.

In this paper we describe our effort in joining the grid environment with the virtual observatory.

## 2. Virtual Observatory Theoretical background

During the development of the Astronomical Virtual Observatory (AVO) project, funded by the EU, the need to access theoretical data within the VO framework was identified to be an important requirement. One of the scenarios call for the possibility of extracting simulated data deriving from models from an archive of theoretical data, or producing the simulations themselves, to allow comparison with observations extracted from VO archives.

The requirements on the VO imposed by this scenario are being analysed within the Design Study 4 of the VO-Tech project, funded by the EU within the Sixth Framework program (FP6). This scenario calls for the concept of "virtual data": simulations or theoretical data, transparently for the user, are either retrieved from an existing data store (e.g. an archive), or built on-the-fly by a computing facility.

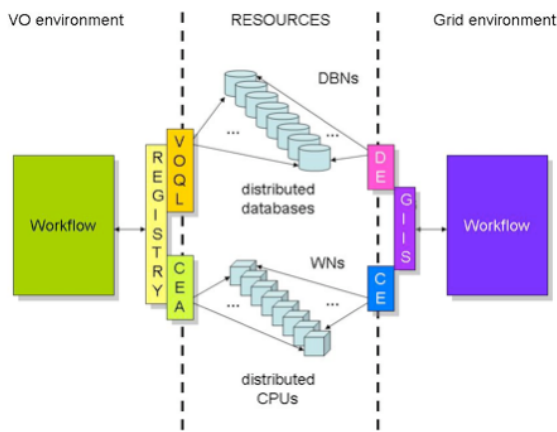

**Figure 1** Two-way infrastructural approach for linking computing resources and data resources, and viceversa, within the VO and Grid environments.

The idea of "virtual data" is that new data, derived from the analysis of raw data, or generated by simulations, should not be discarded after use but saved for retrieval, in the same way that raw data are stored. In a later calculation, the Grid's software could then automatically detect whether a computation had to be started from scratch, or could take a short cut using a previous result stored somewhere in the system.

The "virtual data" concept was developed by the GriPhyN project [7] in the US, and since taken further by the International Virtual Data Grid Laboratory project. In this case, the user doesn't need to know whether a given dataset is generated on-the-fly or read from disk, all s/he is interested in is getting that particular dataset. Of course, issues like comparing the costs and benefits of storing computational results versus generating them when required, etc., should be addressed. A practical use of the "virtual data" concept would be extremely beneficial for the VO community dealing with theoretical data.

The generation on the fly of simulated or theoretical data requires the definition of an infrastructural approach for linking computations with data, and vice versa, within the VO.

Two separate approaches are possible for interconnecting the VO with a computational grid. They depend from the user's choice of starting a workflow from within the VO environment or the Grid one (see Figure 1). This probably depends from the nature of the problem to be solved: more data-oriented or more computation-oriented, respectively.

## 3. Access to VO resources from GRID

A computationally intensive astronomical application running within a Grid environment may need to access data stored in the VO: a grid element capable of accessing databases in a VO-compliant way is therefore needed to be included in the Grid middleware. The implementation of a Data Source Engine (DSE) is the natural extension to allow databases to be accessed within the Grid environment. Moreover a coherent modification of Grid collective resources (Resource Broker, Information Index, etc) needs to be implemented in order to locate and interact a Grid DSE [11].

### 3.1 A Query Element for LCG

In the current Grid architecture, the Resource Framework Layer (RFL), the Information System Framework (IS) and the Information Data Model encompass semantic and interaction models for Grid resources. They are able to retrieve run time data and information from the defined Grid node elements: Computing Elements (CE) and Storage Elements (SE).

However, it exists different types of virtual computing machines such as Java Virtual Machine, Parallel Virtual Machine and Data Source Engine.

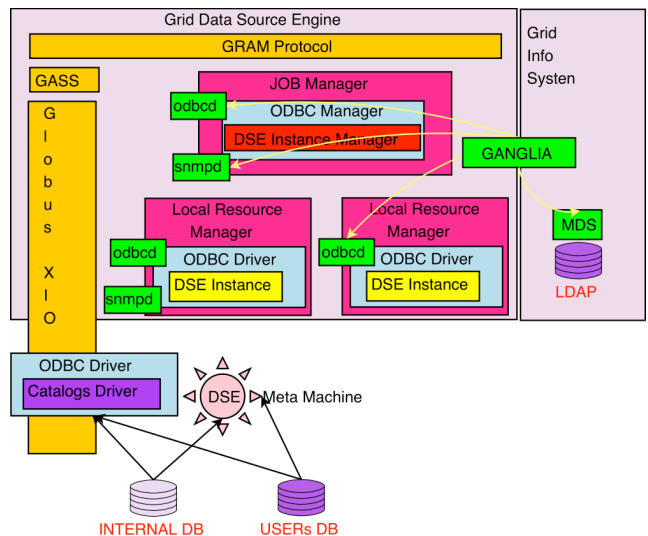

**Figure 2** G-DSE model

In [1] the authors demonstrate that it is possible to define an abstract computational model for Grid and to extend the idea of "job submission" in Grid, to encompass queries evaluation, dynamic java class loading, logic and functional resolution, inference and reasoning performing.

They show that it is possible to implement the interaction between Grid RFL and the specific computing environment (DSE, JVM, and Reasoner) at Grid Fabric layer.

In this paper, we focus on a RDBMS that is a kind of DSE and that we consider an abstraction of classical software computing-machine being modelled by an RDBMS Processor, with a relational instructions Set, and with a RDBMS memory model plus RDBMS I/O Model. On Figure 2, we present an architectural model of the G-DSE as presented by [1].

The deployment of a DSE in Grid implies: the definition of information schema to advertise on the IS of the DSE resources and the extension of the Globus Jobmanager. (see Figure 3).

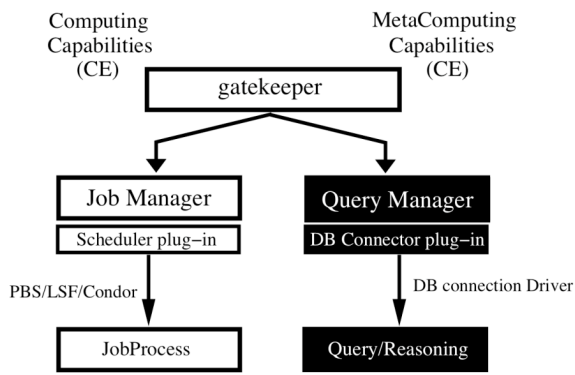

Figure 3 The new DBmanager. A parallel view of the jobmanager and of the dbmanager of globus developed from G-DSE

We deploy on a production Grid a Grid Query Element (QE) that allows the Grid user to interact with a RDMS: creating and administrating a database (DB), doing select queries to get data or using the computational capabilities of a RDMS to perform operation on data.

## 3.2 VO resources in the GRID

The QE and the modified Global Information Service (namely the BDII) and Resource Broker allow locating, querying and managing a generic DB. However, to integrate a VO resource in the Grid, we also extend the information schema with the additional constraint of compliance with the Resource and DataService schemas proposed by IVOA (see Figure 4). Moreover, we develop a specialized Information Provider generating an IVOA compliant set of information. This provider publishes a VO resource in terms of LDAP GlueSchema [2]. A VO to GlueSchema translation library is designed and implemented. When an Astronomical application needs to locate a VO resource in the Grid, it specifies the resource description in terms of IVOA compliant VOQL documents. Those documents are translated into GlueSchema LDAP language to query the Grid BDII. The result of the query is the pointer to any QE with the specified resource.

## 4. CONCLUSION

We present a new approach to bridge the Grid and the VO, based on the Globus QE. The QE and the extended Grid IS allow publishing a VO resource in the production Grid. The Job Description Language and DAG support can be used. The LCG file catalogue can be used to copy, backup and replicate the DB in the Grid allowing multiple access point to the Astronomical resources for parallel DB access or to balance the load of single resources. A grid of DB now is integrated with and within the Grid. The integration of a VO resource in the Grid allows developing applications that require complex workflow that use computing and meta-computing resources (RDMS): data query (eventually requiring the RDMS to preprocess some data), computations and store of the results on a new DB. This can be done using the standard grid tools to build workflow as DAG.

The need to make the relation between these kinds of infrastructure tighter is generated by a whole class of astronomical applications needing massive amounts of computational power: large data mining efforts, production of theoretical data, or simulations to be compared to observational data.

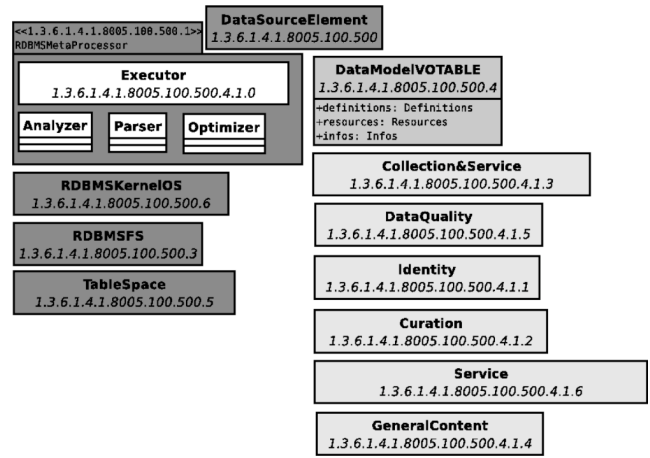

Figure 4 GlueSchema extended for DSE and metadataModel Filesystem.

While a path for allowing this interconnection, the CEA [5], has been sketched already within the VO, its implementation is not mature enough to allow production software to be run on a grid from within a workflow running from within the VO environment. More work is needed in this area. Conversely, if application code is run within a computational Grid environment in order to use the resources available, there is a requirement to access VO-compliant databases to retrieve the data and information needed to complete the computation being made. In this case, the G-DSE is a successful way to join the GRID with VO.

The items discussed here however do not take into account the fundamental issue of security. The authorisation and authentication mechanisms of VO and of the national and international computational grids are bound to be different, and possibly incompatible. It is to be seen if a single sign-on will still be possible, or else some sort of wrapping mechanism can be found to allow minimising the awkwardness of separate log-ins.

It is important that issues of such kind are tackled and solved during the paths of both the VO and the Grids to maturity.